\documentclass{vgtc}                          

\ifpdf
  \pdfoutput=1\relax                   
  \usepackage{graphicx}                
  \DeclareGraphicsExtensions{.pdf,.png,.jpg,.jpeg} 
\else
  \usepackage{graphicx}                
  \DeclareGraphicsExtensions{.eps}     
\fi%

\graphicspath{{figures/}{pictures/}{images/}{./}} 

\usepackage{microtype}                 
\PassOptionsToPackage{warn}{textcomp}  
\usepackage{textcomp}                  
\usepackage{mathptmx}                  
\usepackage{times}                     
\usepackage{cite}                      
\usepackage{tabu}                      
\usepackage{booktabs}                  

\usepackage{soul}
\usepackage{todonotes}
\usepackage{gensymb}

\onlineid{0}

\vgtccategory{Research}

\vgtcinsertpkg

\title{Improving Understanding of Biocide Availability in Facades through Immersive Analytics}

\author{Negar Nouri\thanks{e-mail: negar.nouri@hs-coburg.de} 
\and Snehanjali Kalamkar\thanks{e-mail: snehanjali.kalamkar@hs-coburg.de} %
\and Forouzan Farzinnejad\thanks{e-mail: forouzan.farzinnejad@hs-coburg.de}
\and Verena Biener\thanks{e-mail: verena.biener@hs-coburg.de}
\and Fabian Schick\thanks{e-mail: fabian.schick@stud.hs-coburg.de}
\and Stefan Kalkhof\thanks{e-mail: stefan.kalkhof@hs-coburg.de}
\and Jens Grubert\thanks{e-mail: jens.grubert@hs-coburg.de}
}

\affiliation{\scriptsize Coburg University of Applied Sciences and Arts, Germany}

\teaser{
  \centering
  \includegraphics[width=\linewidth]{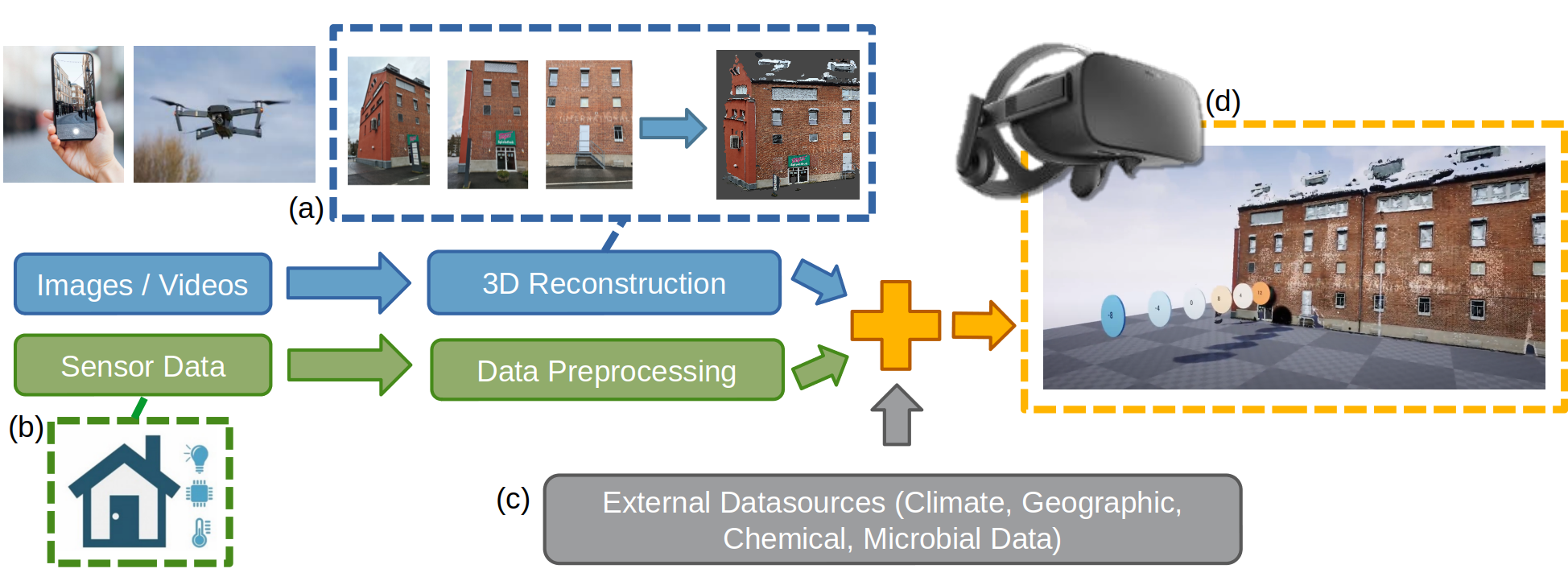}
  \caption{Proposed pipeline: The system consists of four components. (a) Capturing images or videos for 3D reconstruction. (b) Embedding sensor data from the building facade for data processing. (c) Integrating external data sources into the virtual reality environment. (d). Immersive visualization using virtual reality headsets for analyzing and evaluating data.}
  \label{fig:proposed method}
}

\abstract{The durability of facades is heavily affected by multiple factors like microbial growth and weather conditions among others. Biocides are often used to resist these factors and protect the facades. However, the biocides get washed out due to rains and other factors like geometric structure of the facade, orientation of the building. It is therefore, important to understand how these factors affect the durability of facades, leading to a requirement of expert analysis. In this paper, we propose a technical pipeline and a set of interaction techniques to support data analysis within the immersive environment for our case study. Our technical pipeline mainly consists of three steps: 3D reconstruction, embedding sensor data and visualization and interaction techniques. We made a formative evaluation of our prototype to get insights from microbiology, biology and VR experts. The remarks from the experts and the results of the evaluation suggest that an immersive analytic system in our case study could be beneficial for both experts and non-expert users.  %
} 


\begin{document}


\firstsection{Introduction}

\maketitle



Various factors affect the durability of facades, such as microorganisms, material, microclimate, and construction method. Microorganisms can cause greenish or blackish coatings on house facades, resulting in structural damage and high remediation costs. Therefore, biocides are often used to improve the weatherability of facades. However, these are gradually dissolved out by rain, which can reduce their effectiveness as well as endanger the environment. In addition, weathering factors such as humidity, temperature, solar intensity, material properties, wind, dew, etc. influence the extent of degradation or release as well as microbial infestation of facades and thus ultimately the long-term weather resistance. Finally, the geometric structure of the facade and contextual factors such as the orientation of the building can have an influence on the washout of biocides. The joint exploration and analysis of these manifold parameters are complicated. 

Hence, within this paper, we propose to utilize Immersive Analytics to support the exploration of biocides in facades of a building. Specifically, we propose to merge data originating from sensors embedded at the building, additional datasources (e.g., building orientation, location) as well as  3D reconstructions of the building itself into a joint immersive analytics system. 

To this end, we propose both a technical pipeline as well as a set of interaction techniques to support data analysis within the immersive environment.

\section{Related Work}

Our work is based on prior contributions in immersive analytics, 3D reconstruction and sensor data visualization.

Immersive Analytics is a steadily growing field with substantial potential to aid the analytic process of domain experts but also numerous challenges \cite{ens2021grand}. For recent surveys we refer to the works by Fonnet et al. \cite{fonnet2019survey} and Kraus et al. \cite{kraus2022immersive}. Our work is specifically related to immersive and situated analytics systems focusing on analyzing data within physical contexts or 3D reconstructions of physical sites. 
For example, Benko et al. \cite{benko2004collaborative} presented VITA (Visual Interaction Tool for Archaeology), a multimodal, collaborative mixed reality system for documenting and exploring excavation sites using a combination of 3D reconstructions and additional digital data. 
Skarbez et al.\cite{skarbez2019immersive} gave consideration to a number of possibilities, including how 3D scans can be viewed at several scales and how this could change the analytical process for archaeologists. 


While the majority of conventional urban design tools use 2D displays, immersive analytics based on head-mounted displays (HMDs) can be a potential alternative. 
To this end, Zhang et al. \cite{zhang2021urbanvr} presented a comprehensive immersive analytics system that incorporates analysis and visualization methods to assist decision-making in the urban site development. The efficiency and viability of their proposed system have been confirmed by quantitative user studies and qualitative expert feedback.

Sun et al. \cite{sun2019immersive} combined geographical with time-series data from chemical sensors as well as meteorological data in an immersive virtual reality environment. 

Natephra et al. \cite{natephra2019bim} proposed to combine sensor data with building information modeling (BIM) data. The findings demonstrated that the developed system could display various indoor environmental data of the building using VR technology, giving users access to information that can be used to monitor the building’s indoor thermal comfort conditions in real-time. 
The HoloSensor project sought to improve the visualization and visual analytics data sourced from various sensors throughout various locations inside a building \cite{jang2018holosensor}.

On the other hand, the use of data representations arranged in connection to pertinent things, places, and people in the real world for comprehension, sensemaking, and decision-making is known as situated analytics. Situated analytics enables users to easily analyze virtual data that is situated in the real environment while gaining access to the power of the data and analysis \cite{thomas2018situated}. Related to that, Ens and Irani also put out the idea of “Spatial Analytic Interfaces”, which intends to use spatial interaction to facilitate situated analytical activities \cite{ens2016spatial}.

\section{Immersive Analytics System}

The hazard potential of a building facade depends on a large number of structural, microclimatic, and material parameters. For an evaluation of the influencing variables as well as clarification for technically remote users, fast and precise data evaluation as well as interactive data visualizations are of particular importance. 
Typically, microbiology researchers use 2D desktop applications or web dashboards to analyze and evaluate biocide data, such as biocide concentrations in the building facade. However, it is rather complex to investigate the correlation between multivariate data with attributes such as temperature, moisture, and biocide concentration in facades using these tools. 
Further, it can be challenging for experts to visualize this multivariate information using single or multiple physical displays and to represent them with respect to their spatial properties.

Hence, we aim to visualize data from various sensors embedded in the building facade in a virtual reality environment, which aims at supporting experts in comprehensively understanding the effects of multivariate data (such as temperature, moisture, and biocide concentration on the facade of the building) in the spatial context of the facade. 

To this end, we propose an immersive analytics system, consisting of three parts: 3D reconstruction, embedding sensor data, and an accompanying set of 3D visualization and interaction techniques.

\subsection{3D Reconstruction}
 Representing a 3D model of building facades in  virtual reality allows experts to observe multivariate sensor data in the spatial context they originated from. 
 For example, the 3D model can support domain experts to gain a better understanding of the architectural design and the  position of  doors and windows, which might affect measurements. Furthermore, they will also have the opportunity to observe variation and correlation of the measurements in different areas of the facade.


Photogrammetry 
is a mature 3D reconstruction approach that also can be utilized with ordinary smartphones, as it only requires a set of ordinary images. Recent developments in Structure from Motion (SfM) and Multi-View Stereo (MVS) have enabled the automation of some phases, such as camera motion determination and scene geometry reconstruction. Still, achieving satisfying 3D reconstructions, specifically in large-scale environments, can be challenging due to the large number of required images as well as the parametrization of SfM pipelines.

\begin{figure}[t]
	\centering 
	\includegraphics[width=1\columnwidth]{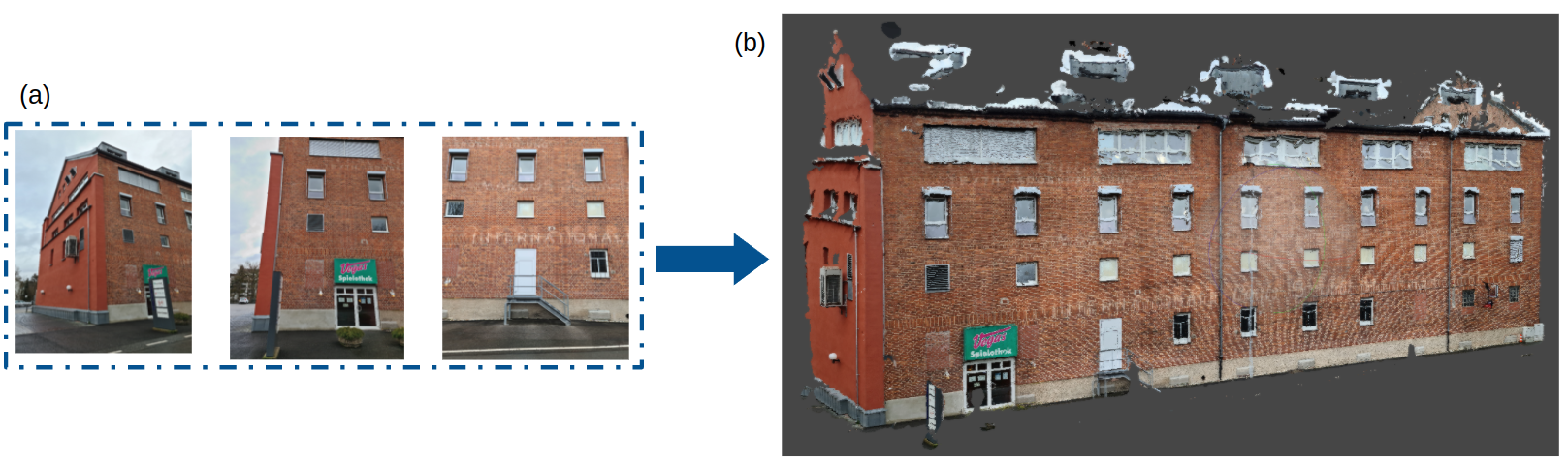}
	\caption{3D Reconstruction: (a) Some sample images of the building facade, (b) 3D model of the building facade.    }
	\label{fig:3Dreconstruction}
\end{figure}

 We employ an SfM pipeline for 3D reconstruction, which is aimed at non-professional users \cite{dietz2021towards}. This pipeline also facilitates the creation of 3D models of the required buildings. Further, for a more comprehensive understanding of the measurements and correlations between variables such as temperature and moisture, the position of the embedded sensor can be integrated into the 3D model using game engine software such as Unreal Engine \footnote{https:\/\/www.unrealengine.com}. Besides, in our case, we used a typical smartphone to capture images and several sample images can be seen in \autoref{fig:3Dreconstruction} along with a 3D model of the building. A complimentary high-quality reference model can be generated by a LiDAR scanner.
 
\subsection{Embedding Sensor Data}

\begin{figure}[t]
	\centering 
	\includegraphics[width=1\columnwidth]{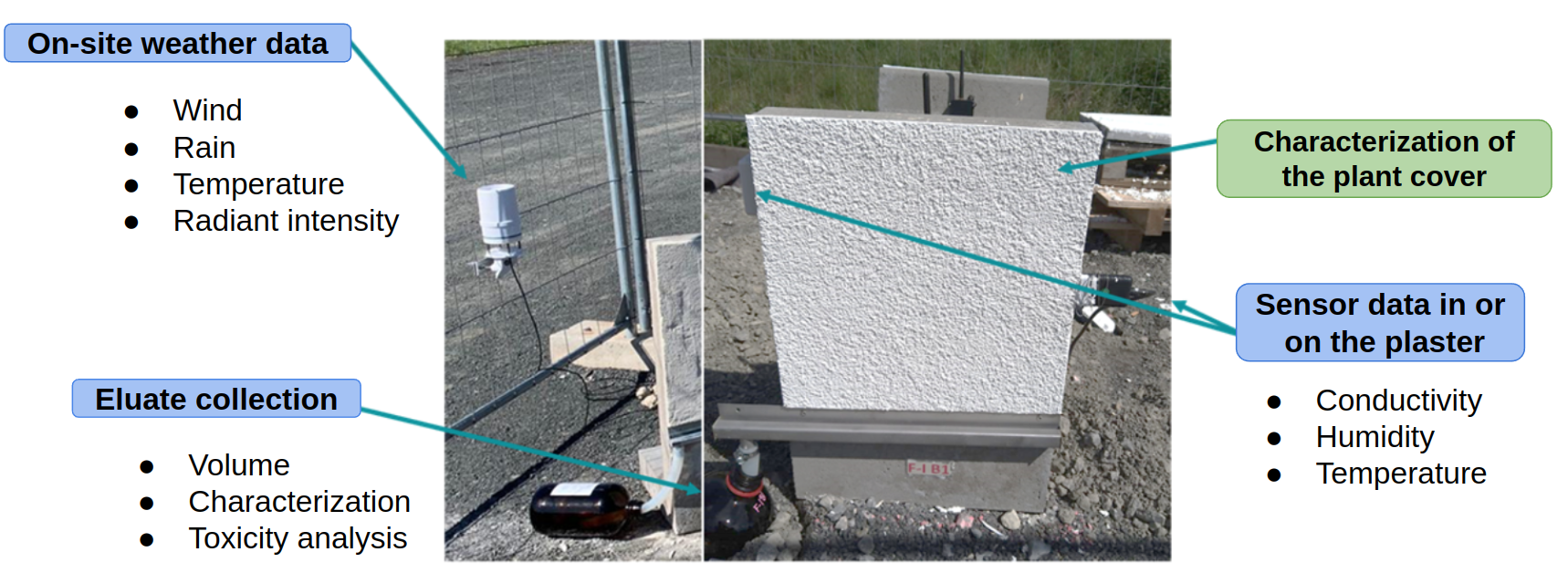}
	\caption{Embedded sensors in the sample building facade and values measured in the actual field}
	\label{fig:iotdata}
\end{figure}

A number of factors affect the durability of facades, such as microorganisms, material, microclimate, and construction methods. In addition, weathering factors such as humidity, temperature, solar intensity, material properties, wind, dew influence the extent of degradation or release, as well as the microbial infestation of facades has an impact on the long-term weather resistance \cite{reiss2021application}. Hence, various sensors need to be employed to for collecting relevant data at or near a facade.  
The three main groups of relevant data include weather station data on-site, data from sensors embedded in the plaster as well as data collected from eluates. In our project, sensors have been embedded in different regions of a sample facade to jointly collect this data (see \autoref{fig:iotdata}).

 A variety of weather data is collected on-site, such as wind, rain, temperature, and radiant intensity. A plaster sensor measures conductivity, humidity, and temperature.  Lastly, eluates collection includes volume, characterization, and toxicity analyses. 
 It is also possible to investigate the characterization of the plant cover. We used the ThingsBoard IoT platform \footnote{https:\/\/www.thingsboard.io/} to store this data.

\subsection{Visualization and Interaction Techniques}



We propose to use three main strategies that support analytical reasoning  of live  data generated from the sensors embedded in the facades. These strategies include time series visualization, data interpolation, and multi-scale navigation.

\subsubsection{Time Series Visualization}

Time series data consists of a set of attributes that change over time and can be categorized into a variety of groups: linear time, cycle time, time intervals, time point, sequential time, etc\cite{fang2020survey}. Sensor data can be measured in different intervals depending on the purpose. Temperature and moisture attributes, for instance, can be measured either monthly or seasonally, which can be referred to as cycle time. Time series data is specifically relevant for our use case as reasoning about biocides in facades needs to take into account changes over potentially many weeks, months or even years. In the case of biocide concentration, we can use a seasonally time series because it will change significantly throughout the different seasons and will assist the expert to analyze the data more effectively. Further, multiple variables can be considered for selecting appropriate time series, such as the correlation between moisture, temperature and biocide concentration, architecture, location, and orientation (e.g., north or south facing) of the building facade, as well as conditions of the area (e.g., if lakes or woods are nearby a facade). 

In principle, various visualization methods can be used for showing different types of time series data (e.g, spiral diagram, calendar view,  ThemeRiver view, and dynamic visualization~\cite{fang2020survey}). In our use case, time series data of sensors is displayed on a spatial referent (the facade). Hence, besides taking the nature of attribute data and the correlation between multiple variables into account, we also need to consider the spatial reference frame for choosing appropriate time-series visualization techniques. 


It is important to consider several factors when displaying time series data in conjunction with a physical referent. A given measurement can be designed with geometric primitives or more complex geometric shapes(e.g., cubes, cylinders, drops of water), depending on the attribute. In our case, cubes or circles can be used to display temperature data. 
A cube for example, can be used to show multiple values related to the data point on it's different faces. We used diverging colormaps to visualize our attributes, temperature and moisture. Specifically, for temperature, all temperature values above 0\degree\ were interpolated with RGB values towards a yellow to red color space, indicating more heat. All temperature values below 0\degree\ were interpolated with RGB values towards a blue color space, indicating less heat.



\begin{figure}[t].
	\centering 
	\includegraphics[width=1\columnwidth]{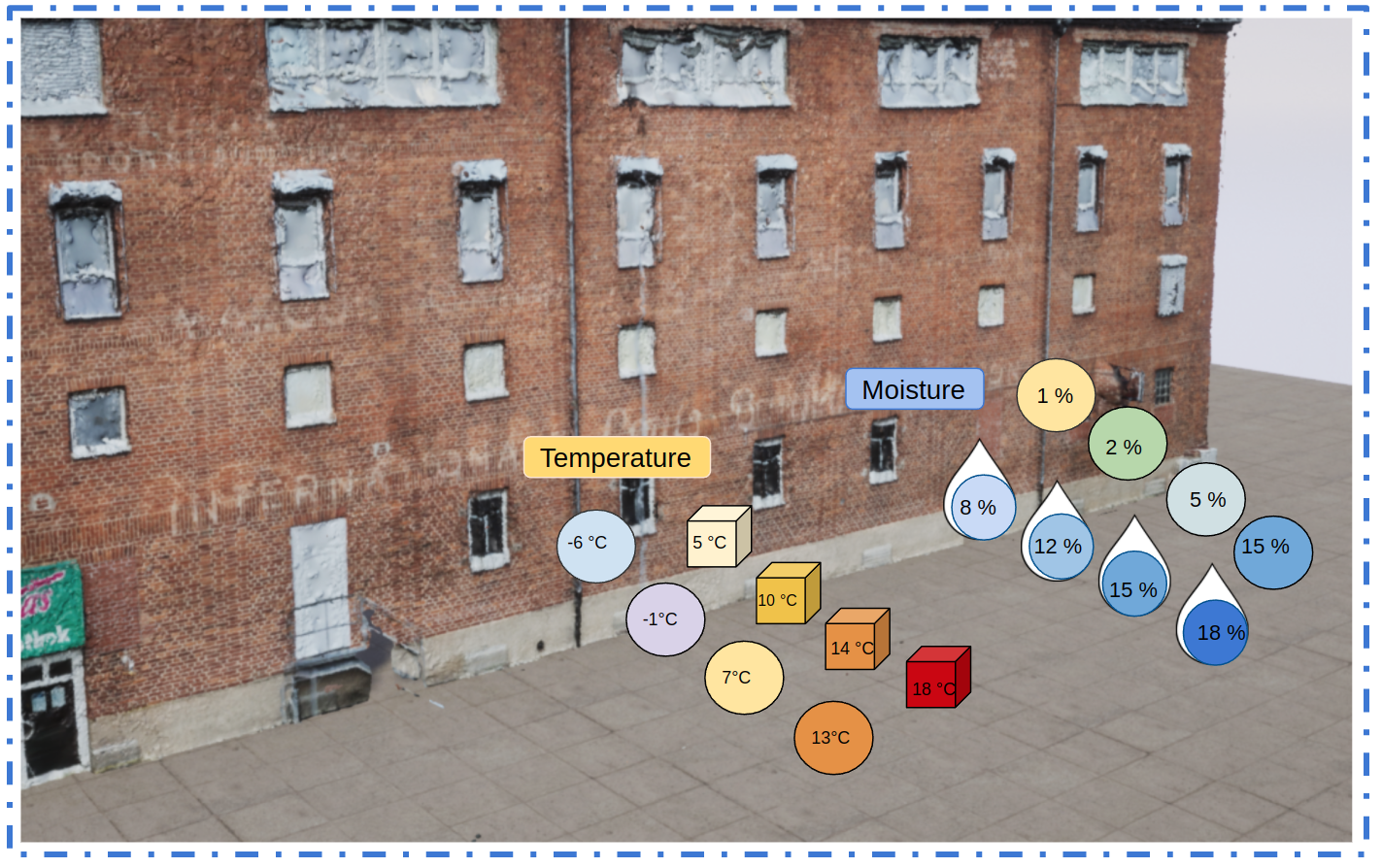}
	\caption{Time series visualization: Temperature and moisture have been shown as an example of the sensor data that vary over time.}
	\label{fig:Timeseriesvisulaization}
\end{figure}

Water droplets and circles are used to display moisture data, and they can also be displayed seasonally (see \autoref{fig:Timeseriesvisulaization}). Similar to temperature, all moisture values above 50\% were interpolated with RGB values towards a blue shade, indicating higher humidity. All temperature values below 50\% degrees were interpolated with RGB values towards a yellow shade, indicating a dry weather.

\subsubsection{Data Interpolation}

Since current measurements of the sensor data are limited to specific positions at the facade of the building, but data on further locations or whole areas of the facade can be relevant for facilitating understanding, it is necessary to perform attribute interpolation. 
However, simple  interpolation schemes like linear interpolation are not suitable for all measurements. 
For example, in the case of moisture data, the design of the building facade and the position of the doors and windows will influence moisture distribution.

Generally, spatial interpolation can be subdivided into point and areal interpolation. In areal interpolation, attribute information is transferred from source zones with known values to other regions.  Point interpolation predicts values at unknown locations based on information from sample points, which are considered to vary over space continually. Point interpolation approaches include inverse distance weighting, ordinary kriging models, kriging with external drift, and spline \cite{comber2019spatial}. However, spatial interpolation profoundly impacts a variety of aspects, including sample size, sample design, and the nature of the data. Furthermore, it is complicated to predict how all those factors impact the performance of the interpolation method \cite{srivastava2019gis}.  

\begin{figure}[t]
	\centering 
	\includegraphics[width=1\columnwidth]{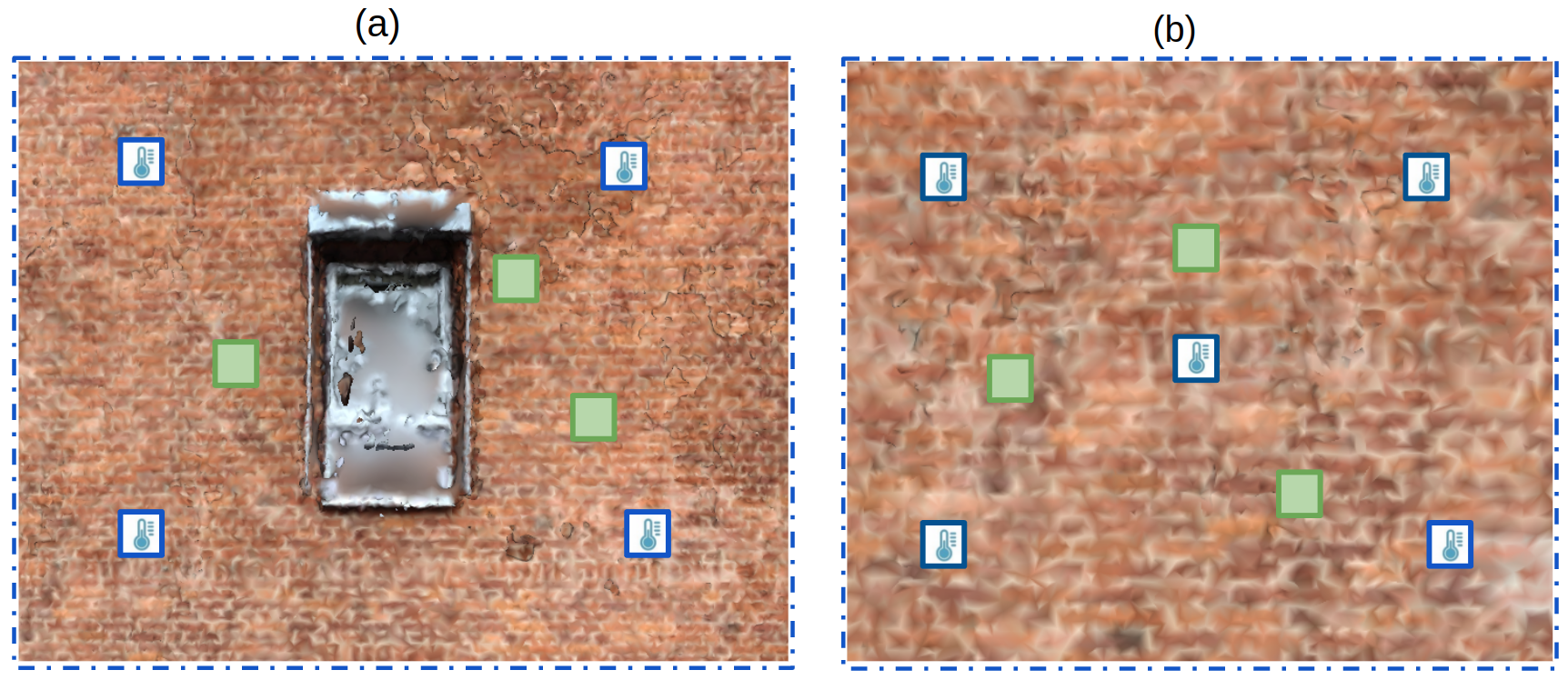}
	\caption{Data interpolation: white rectangles are the position of embedded sensors on the building of the facade and the green rectangles are the positions where data should be interpolated. (a) the building facade with window, (b) the building facade without window.}
	\label{fig:Datainterpolation}
\end{figure}

In our current interpolation, for instance, in \autoref{fig:Datainterpolation} (b), point-based interpolation (kriging) is employed to determine values in unknown areas such as green rectangle. 
Please note, that this assumes homogenous regions of the facade. However, this might not always lead to corrent values.
For example, as can be seen in \autoref{fig:Datainterpolation} (a), in the given area of the facade, which includes embedded sensors and windows, it is challenging to apply interpolation to calculate data in an unknown point (e.g., green rectangle) due to the presence of the windows (that constitute areas without biocide concentration). Besides, other factors such as sample size, the type of embedded sensor, and building material can influence spatial interpolation efficiency. Hence, it can be challenging to select an appropriate spatial interpolation method for a given specific input data set due to variation within the data. In our case, one possible alternative to overcome the problem of the existence of the windows is to embed more sensor data close to the windows, which will help  to apply a more accurate interpolation between all those sensor data.

 Further, multiple factor analysis can be performed among different sensor data, as discussed in embedding sensor data. For example, moisture and temperature can influence biocide concentration among other factors. Particular parts of the facade of the building, such as windows and doors, can absorb more water during rainfall. Accordingly, the sample area will record different values for moisture, temperature, and biocide construction.  In \autoref{fig:Datainterpolation}, sample area (a) has more moisture and less temperature compared with sample area (b) and hence, has less biocide concentration and more microbial growth. Consequently, moisture and temperature can be used to predict biocide concentrations and microbial growth in different areas of the facade. Please note, that our aim is not to propose novel algorithms for multifactor analysis, but to enable domain experts to investigate multiple factors in the spatial context they arise.



\subsubsection{Multi-Scale Visualization}


Scale change in virtual reality has been investigated for various applications that allow the user to view the diverse elements of the virtual environment and become more familiar with its intrinsic hierarchical patterns~\cite{abtahi2019m}. 

\begin{figure}[t]
	\centering 
	\includegraphics[width=1\columnwidth]{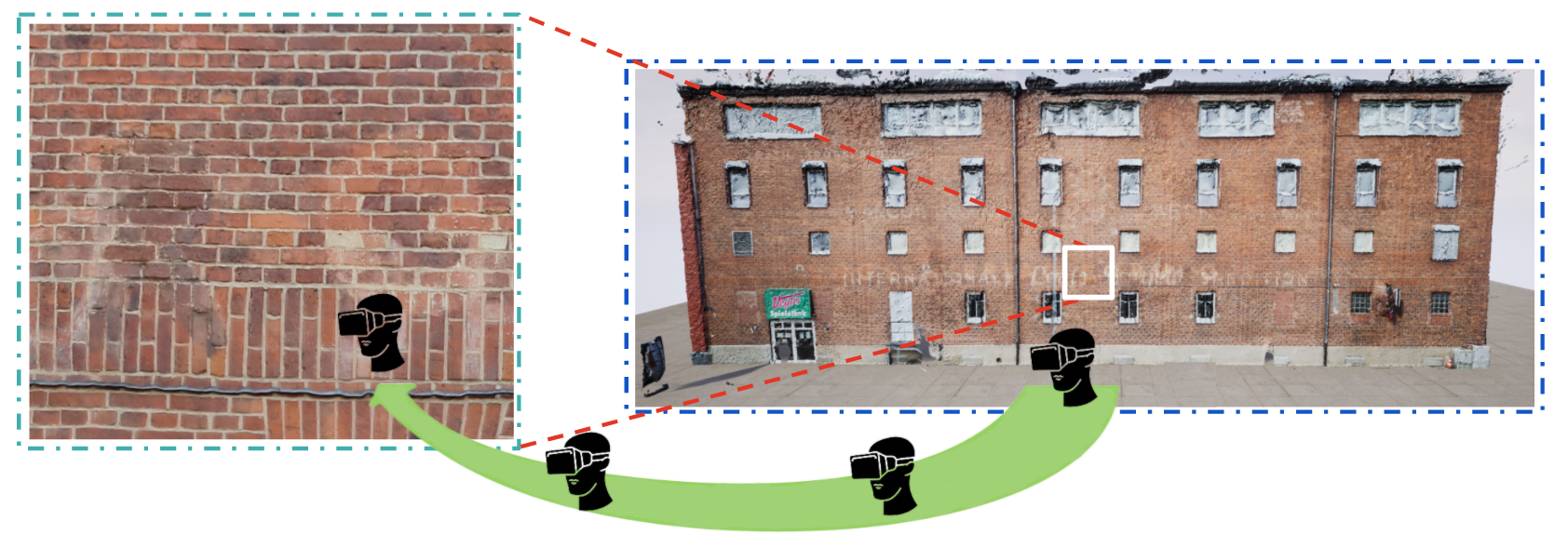}
	\caption{Multi scale navigation: In zoom-in transition techniques, the user's pose is animated, and the view is scaled in two different scales, such as full view and high-resolution view }
	\label{fig:multiscale}
\end{figure}

In the context of analyzing biocides on facades multiple scales are relevant. For example, the domain expert might want to get an overview about selected data distributions on the whole facade as well as regions around specific sensors (see \autoref{fig:multiscale}). 
In our use case, as seen in \autoref{fig:multiscale}, right, initially, the user and the environment are on the same scale (i.e. as a human would observe the facade if she is standing in front of the physical building). While manifold travel techniques have been proposed (for an overview we refer to \cite{al2018virtual, di2021locomotion}), in our current implementation, we enable both steering-based travel (via the D-Pad of a controller) in the initial scale and teleporting (both in the initial and into/out of further scales).  
Various methods including orbit-then-zoom, curvilinear zoom, and zoom-then-rotate can be utilized \cite{lee2022comparison} for transitioning between scales. In our current implementation we use a discrete scale change (i.e. teleporting to a new location while rescaling the underlying 3D model). 


One specific aspect that needs to be taken into account when employing multi scale navigation within a 3D reconstructed environment is the scale of the original input images and the generated textures of the reconstructed 3D model. To enable meaningful multi scale navigation we capture the facade at multiple scales. An initial set of images is taken from the point of view of an observer standing in front of the building. A second set of images is taken at a closer range, but only for potentially relevant regions, such as around sensors. For rendering, all images can be utilized to generate a single 3D model with high and low resolution areas. Still, this model is bound by the maximum texture size that the 3D reconstruction system supports. As an alternative, multiple 3D models can be reconstructed and switched during runtime (e.g., using level of detail methods~\cite{luebke2003level}).

\label{sec:conclusion and Future work}





\section{Implementation}
We have implemented an initial immersive analytical system, used to visualize 
the data from sensors embedded in the physical environment. This involved the following steps: 3D reconstruction of the required physical environments, receive sensor data and analyzed data for required attributes, and 3D visualization of respective data points relative to the environment in VR (see \autoref{fig:proposed method}).
 

To accomplish 3D reconstruction, we utilize an open-source pipeline that can be utilized by non-professional users\cite{dietz2021towards}. 
The following steps are employed in 3D reconstruction. First, the user uploads a video whose frames will be used for 3D reconstruction. There are some requirements that need to be considered for the video quality to give the best result, such as sharpness, exposure, and noise. Those factors are checked by the web-based system, and the user is notified about the video quality. Next, the frames are extracted from the video and filtered based on blur detection. The user is then notified regarding the usability of the frames. Lastly, the 3D modeling process begins and the user can download the 3D model as a ZIP file \cite{dietz2021towards}. Alternatively, existing SfM systems which require a deeper understanding of the underlying processes by the users, like AliceVision \cite{alicevision2021} can be employed.

In order to perform data augmentation and analysis, the first step is to acquire the available data. We used the pre-recorded field data, available on a custom ThingsBoard IoT server. The different data sources that we used were namely air temperature, humidity, wall temperature, wall humidity, and conductivity. 
We fetched the weather station data and recorded sensor data into Unreal Engine via the VaRest plugin using REST API calls to ThingsBoard IoT platform.

Next, the 3D models and the processed sensor data were transformed into interactive 3D visualizations and made accessible in a virtual reality environment. 
Additionally, this environment should also enable the cross-regional comparison of multiple objects with similar properties (e.g. altitude, weather conditions). For this purpose, open data sources (e.g. www.openstreetmap.de for position data, opendata.dwd.de for climate data) will be connected. Within the project, analytical measurement data, climatic data, material parameters, and infestation forecasts are correlated to overlay these data with building data.

\begin{figure}[t]
	\centering 
	\includegraphics[width=1\columnwidth]{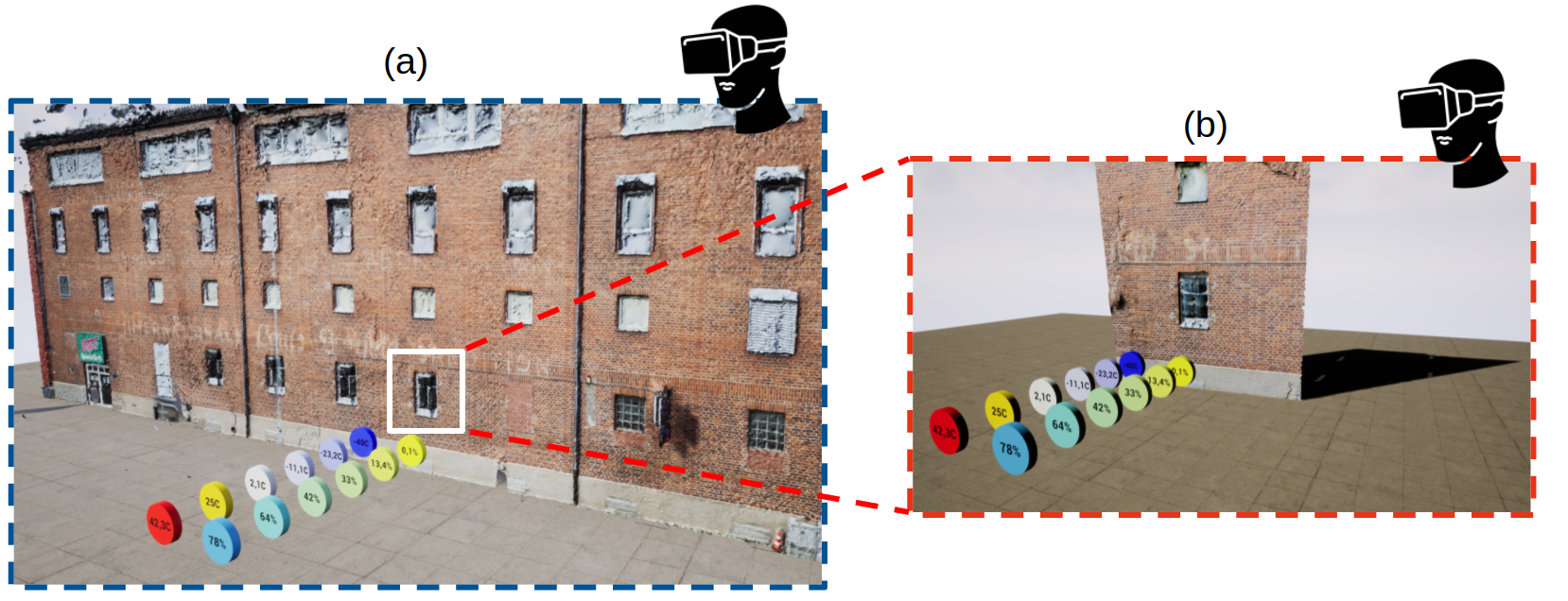}
	\caption{Implementation: The first prototype of our immersive analytics system. (a) 3D model of the building facade along with temperature and moisture measurements in a time series (b) A sample area of the facade }
	\label{fig:sketchgraphs}
\end{figure}

The initial prototype was implemented using Unreal Engine 4.27.
The scene consists of a 3D model of a building that represents the building in that the sensors are embedded.
In front of the buildings facade, the 3D objects representing the sensor data points in a time series are displayed (see \autoref{fig:sketchgraphs}).
The different measures can be shown or hidden by pressing the corresponding buttons on a controller. We used left controller button to visualize temperature and the right controller button for moisture. Similarly, mappings for more attributes can be made in the VR environment. We used the trigger button to switch between the  areas of interest, this switched the view from complete building facade along with temperature and moisture measurements (see \autoref{fig:sketchgraphs} (a)) to the sample area of the facade (see \autoref{fig:sketchgraphs} (b)).

%

\section{Formative Evaluation}
\begin{figure}[t]
	\centering 
	\includegraphics[width=1\columnwidth]{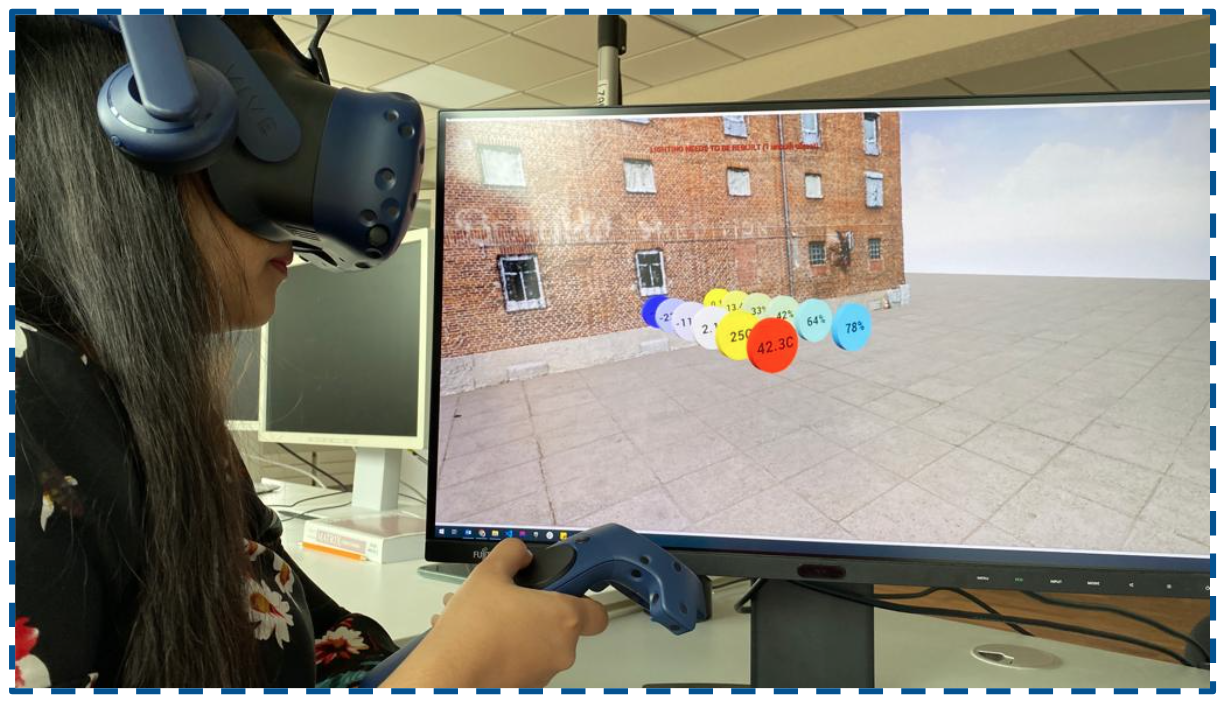}
	\caption{Expert evaluating the first prototype of our immersive analytic system}
	\label{fig:/evaluation-from-VR-expert}
\end{figure}

We conducted an initial formative evaluation with domain experts, which allowed us to gain valuable information on how to improve our system. We demonstrated the system to one VR and two microbiology experts (2 male, 1 female, mean age = 27) and interviewed them about their observations in the VR environment. 
The experts were employees or researchers at a university. All experts had normal, or corrected to normal eyesight and they saw everything clearly in the virtual environment. The two microbiology experts had no previous experience with VR. All experts were informed about the procedure and the content of the study, signed a consent form and filled out a demographic questionnaire. Next, the microbiology experts received a short introduction using the HMD (an HTC Vive Pro Eye) and how to work with the associated controller (see \autoref{fig:/evaluation-from-VR-expert}).

Next, while the experts were wearing a VR HMD and testing the application, they were interviewed. They were asked about how they felt about the application, what they liked or disliked, how practical it was, what kind of function they would like the system to have, which problems occurred, and what they would improve. Biology experts were also asked  if they could imagine using VR for data analysis in the future. At the end, we gained valuable information from the experts through the interview. 


In the beginning, we asked the biology experts how they felt while observing the system. Both (P02 and P03) liked working in virtual reality. They preferred the teleportation-based navigation method over the steering-based one. 
In addition, regarding 3D models, a more detailed model  was suggested. They also liked how we presented temperature and moisture over time. 
Nevertheless, P02 pointed out that for the moisture element, a different color scheme can be used (e.g., moisture measurement: zero percentage can be red). A switching button was also recommended to attach to the sensor data, which enables the expert to select sensor data for daily, monthly or seasonal observation based on their goal. Both experts also suggested using a heat map instead of displaying temperature attributes in different time series. In addition, P03 stated that a heat map would help them gain information quickly; for example, red indicates regions with high temperatures. However, in that case, displaying information in heat map layers can be achieved by using data interpolation, which can vary over time by different layers. With respect to visualizing biocide concentration, P02 stated that a layer could be mapped to the wall so that it can be thick in the region with more biocide concentration and thin with less biocide. For visualizing sensor data,  they were inquired to name suitable places for the sensors to embed in the facade. They pointed out it depends on the type of sensor; for instance, the moisture sensor should be placed in the area close to windows, walls, and the plumbing, which creates a moist surrounding on the facade in case of the rainy season.


In the end, the biology experts pointed out that they could imagine using VR for analyzing data in the future if some conditions are met, such as having lighter HMDs with higher resolution. 
Besides, they said this system could benefit non-expert users, such as building owners and related companies interested in investigating biocide's impact on building facades.

The VR expert also proposes the use of zoom or scroll feature to display temperature and moisture information based on our target. Scrolling or swiping is recommended to see all the elements over time. Also, the expert indicated that it would be beneficial to have information move through time. 
For example, initially, the temperature data points of the first four months could be displayed, on scrolling the data points from the next four months could be shown and so on. In order to apply the multi-scale navigation technique, the expert suggested that a square around the sample area could be used so that the observer could point through it. For displaying the sample area, it is recommended that the rest of the building is still visible but might be blurred. For zooming, an animation was suggested that moves the user closer. The expert speculated that analyzing data in a VR environment could be more effective than analyzing data in a computer since a VR environment could help to observe which data belongs to what part of the building and see the differences at one glance. However, using a computer could still be preferable for calculating statistical measurements (e.g., mean, median, etc.). Furthermore, the VR expert suggested a feature allowing the user to easily add their their own buildings and sensor information.

After the immersive experience all three experts filled out a short questionnaire.
This included three individual questions with a seven-item-likert scale with 1 meaning "not at all" and 7 meaning "very much". 
The questions "I would find the application to be useful" and "I would find the application easy to use" where rated with two 6 and one 7 ($m=6.3$). The question I would have fun interacting with the application" even got two 7 ($m=6.67$). This indicates that all experts have a very positive attitude towards such a system.
The simulator sickness questionnaire \cite{kennedy1993simulator} resulted in a mean of $18.7$ ($sd=9.9$) which is rather high, but could be explained by the microbiology experts having no prior VR experience \cite{chang2020virtual} and the very early state of the prototype.

\section{Discussion, Conclusion, and Future Work}

We presented an initial immersive analytics prototype for supporting domain experts in understanding biocides in and microbial infestation of facades. To this end, we propose to combine 3D reconstruction of physical buildings with sensors embedded in the facades and to make these data jointly available in an interactive VR system. We furthermore proposed three important visualization and interaction aspects of this immersive analytics system (time series visualization, data interpolation, and multi-scale navigation) and conducted an initial formative evaluation with domain experts. 

Our system initially focuses on visualizing only a few selected sensors. 
One challenge is how to not only visualize individual data sources but how to develop meaningful multivariate visualizations that add meaning within the context of the spatial referent (the facade).
 Further, data can come at different levels of granularity, which again  could be spatial or temporal in nature. 

 With respect to the 3D reconstruction a number of factors need to be taken into account (such as the level of expertise of the people on site and the need for capturing images at multiple scales). In addition, it is difficult to reconstruct some reflective, transparent, and non-rigid objects due to the inherent problems of image-based 3D reconstruction with textureless surfaces.
 
 

This project can be improved in several ways in the future. Developing visualizations for physical locations of the sensor in the 3D world, such as DNA structures of microbes found at the sample site,  biocide
concentration as 3D bar charts with respect to time, and temperature
and moisture measurements with respect to time.   
Finally, biology experts pointed out the amount of biocides being still present within the facade material can be predicted based on material parameters, starting concentration and weathering data which can need to be considered in future.

\section*{Acknowledgements}
This project received a grant from Bavarian State Ministry for Science and Art, program to promote applied research and development at universities of applied sciences. We would like to thank the Bavarian State Ministry for Science and Art and Coburg University of Applied Sciences and Arts for their support.


\bibliographystyle{abbrv-doi}

\bibliography{template}

\begin{thebibliography}{10}

\bibitem{abtahi2019m}
P.~Abtahi, M.~Gonzalez-Franco, E.~Ofek, and A.~Steed.
\newblock I'm a giant: Walking in large virtual environments at high speed
  gains.
\newblock In {\em Proceedings of the 2019 CHI Conference on Human Factors in
  Computing Systems}, pp. 1--13, 2019.

\bibitem{al2018virtual}
M.~Al~Zayer, P.~MacNeilage, and E.~Folmer.
\newblock Virtual locomotion: a survey.
\newblock {\em IEEE transactions on visualization and computer graphics},
  26(6):2315--2334, 2018.

\bibitem{benko2004collaborative}
H.~Benko, E.~W. Ishak, and S.~Feiner.
\newblock Collaborative mixed reality visualization of an archaeological
  excavation.
\newblock In {\em Third IEEE and ACM International Symposium on Mixed and
  Augmented Reality}, pp. 132--140. IEEE, 2004.

\bibitem{chang2020virtual}
E.~Chang, H.~T. Kim, and B.~Yoo.
\newblock Virtual reality sickness: a review of causes and measurements.
\newblock {\em International Journal of Human--Computer Interaction},
  36(17):1658--1682, 2020.

\bibitem{comber2019spatial}
A.~Comber and W.~Zeng.
\newblock Spatial interpolation using areal features: A review of methods and
  opportunities using new forms of data with coded illustrations.
\newblock {\em Geography Compass}, 13(10):e12465, 2019.

\bibitem{di2021locomotion}
M.~Di~Luca, H.~Seifi, S.~Egan, and M.~Gonzalez-Franco.
\newblock Locomotion vault: the extra mile in analyzing vr locomotion
  techniques.
\newblock In {\em Proceedings of the 2021 CHI Conference on Human Factors in
  Computing Systems}, pp. 1--10, 2021.

\bibitem{dietz2021towards}
O.~Dietz and J.~Grubert.
\newblock Towards open-source web-based 3d reconstruction for
  non-professionals.

\bibitem{ens2021grand}
B.~Ens, B.~Bach, M.~Cordeil, U.~Engelke, M.~Serrano, W.~Willett, A.~Prouzeau,
  C.~Anthes, W.~B{\"u}schel, C.~Dunne, et~al.
\newblock Grand challenges in immersive analytics.
\newblock In {\em Proceedings of the 2021 CHI Conference on Human Factors in
  Computing Systems}, pp. 1--17, 2021.

\bibitem{ens2016spatial}
B.~Ens and P.~Irani.
\newblock Spatial analytic interfaces: Spatial user interfaces for in situ
  visual analytics.
\newblock {\em IEEE computer graphics and applications}, 37(2):66--79, 2016.

\bibitem{fang2020survey}
Y.~Fang, H.~Xu, and J.~Jiang.
\newblock A survey of time series data visualization research.
\newblock In {\em IOP Conference Series: Materials Science and Engineering},
  vol. 782, p. 022013. IOP Publishing, 2020.

\bibitem{fonnet2019survey}
A.~Fonnet and Y.~Prie.
\newblock Survey of immersive analytics.
\newblock {\em IEEE transactions on visualization and computer graphics},
  27(3):2101--2122, 2019.

\bibitem{alicevision2021}
C.~Griwodz, S.~Gasparini, L.~Calvet, P.~Gurdjos, F.~Castan, B.~Maujean, G.~D.
  Lillo, and Y.~Lanthony.
\newblock {A}licevision {M}eshroom: An open-source {3D} reconstruction
  pipeline.
\newblock In {\em Proceedings of the 12th ACM Multimedia Systems Conference -
  {MMSys '21}}. ACM Press, 2021. doi: {{%
10\hspace{.1pt}\discretionary{.}{%
}{.}\hspace{.4pt}1145\discretionary{/}{%
}{/}3458305\hspace{.1pt}\discretionary{.}{%
}{.}\hspace{.4pt}3478443}}


\bibitem{jang2018holosensor}
J.~Jang and T.~Bednarz.
\newblock Holosensor for smart home, health, entertainment.
\newblock In {\em ACM SIGGRAPH 2018 Appy Hour}, pp. 1--2. 2018.

\bibitem{kennedy1993simulator}
R.~S. Kennedy, N.~E. Lane, K.~S. Berbaum, and M.~G. Lilienthal.
\newblock Simulator sickness questionnaire: An enhanced method for quantifying
  simulator sickness.
\newblock {\em The international journal of aviation psychology},
  3(3):203--220, 1993.

\bibitem{kraus2022immersive}
M.~Kraus, J.~Fuchs, B.~Sommer, K.~Klein, U.~Engelke, D.~Keim, and F.~Schreiber.
\newblock Immersive analytics with abstract 3d visualizations: A survey.
\newblock In {\em Computer Graphics Forum}, vol.~41, pp. 201--229. Wiley Online
  Library, 2022.

\bibitem{lee2022comparison}
J.-I. Lee, P.~Asente, and W.~Stuerzlinger.
\newblock A comparison of zoom-in transition methods for multiscale vr.
\newblock In {\em ACM SIGGRAPH 2022 Posters}, pp. 1--2. 2022.

\bibitem{luebke2003level}
D.~Luebke, M.~Reddy, J.~D. Cohen, A.~Varshney, B.~Watson, and R.~Huebner.
\newblock {\em Level of detail for 3D graphics}.
\newblock Morgan Kaufmann, 2003.

\bibitem{natephra2019bim}
W.~Natephra and A.~Motamedi.
\newblock Bim-based live sensor data visualization using virtual reality for
  monitoring indoor conditions.
\newblock 2019.

\bibitem{reiss2021application}
F.~Rei{\ss}, N.~Kiefer, M.~Noll, and S.~Kalkhof.
\newblock Application, release, ecotoxicological assessment of biocide in
  building materials and its soil microbial response.
\newblock {\em Ecotoxicology and Environmental Safety}, 224:112707, 2021.

\bibitem{skarbez2019immersive}
R.~Skarbez, N.~F. Polys, J.~T. Ogle, C.~North, and D.~A. Bowman.
\newblock Immersive analytics: Theory and research agenda.
\newblock {\em Frontiers in Robotics and AI}, 6:82, 2019.

\bibitem{srivastava2019gis}
P.~K. Srivastava, P.~C. Pandey, G.~P. Petropoulos, N.~N. Kourgialas, V.~Pandey,
  and U.~Singh.
\newblock Gis and remote sensing aided information for soil moisture
  estimation: A comparative study of interpolation techniques.
\newblock {\em Resources}, 8(2):70, 2019.

\bibitem{sun2019immersive}
B.~Sun, A.~Fritz, and W.~Xu.
\newblock An immersive visual analytics platform for multidimensional dataset.
\newblock In {\em 2019 IEEE/ACIS 18th International Conference on Computer and
  Information Science (ICIS)}, pp. 24--29. IEEE, 2019.

\bibitem{thomas2018situated}
B.~H. Thomas, G.~F. Welch, P.~Dragicevic, N.~Elmqvist, P.~Irani, Y.~Jansen,
  D.~Schmalstieg, A.~Tabard, N.~A. ElSayed, R.~T. Smith, et~al.
\newblock Situated analytics.
\newblock {\em Immersive analytics}, 11190:185--220, 2018.

\bibitem{zhang2021urbanvr}
C.~Zhang, W.~Zeng, and L.~Liu.
\newblock Urbanvr: An immersive analytics system for context-aware urban
  design.
\newblock {\em Computers \& Graphics}, 99:128--138, 2021.

\end{thebibliography}
\end{document}